\documentclass[preprint, superscriptaddress, showpacs,preprintnumbers,amsmath,amss]{revtex4}
\usepackage{graphicx}

\begin{document}

\thispagestyle{empty}

\title{Theory of the Casimir interaction for
 graphene-coated substrates using the polarization tensor and
comparison with experiment }

\author{
G.~L.~Klimchitskaya}
\affiliation{Central Astronomical Observatory at Pulkovo of the Russian Academy of Sciences, St.Petersburg, 196140, Russia}
\affiliation{Institute of Physics, Nanotechnology and
Telecommunications, St.Petersburg State
Polytechnical University, St.Petersburg, 195251, Russia}
\author{
 U.~Mohideen}
\affiliation{Department of Physics and Astronomy, University of California, Riverside, California 92521, USA}
\author{
V.~M.~Mostepanenko}
\affiliation{Central Astronomical Observatory at Pulkovo of the Russian Academy of Sciences, St.Petersburg, 196140, Russia}
\affiliation{Institute of Physics, Nanotechnology and
Telecommunications, St.Petersburg State
Polytechnical University, St.Petersburg, 195251, Russia}

\begin{abstract}
We propose a theory of the thermal Casimir interaction for
multilayered test bodies coated with a
graphene sheet. The reflection coefficients on such
structures are expressed in terms of the components of the
polarization tensor  and the dielectric permittivities of
material layers. The developed theory is applied to calculate
the gradient of the Casimir force between an Au-coated sphere
and a graphene sheet deposited on a SiO${}_2$ film covering
a Si plate, which is the configuration of a recent experiment
performed by means of a dynamic atomic force microscope. The
theoretical results are found to be in very good agreement
with the experimental data. We thus confirm
that graphene influences the Casimir interaction and can be
used for tailoring the force magnitude in nanostructures.
\end{abstract}
\pacs{12.20.Ds, 78.67.Wj, 65.80.Ck, 12.20.Fv}

\maketitle

\section{Introduction}

Graphene is a 2D sheet of carbon atoms which finds diverse
applications in nanotechnology and other fields due to its
unusual electrical, mechanical and optical properties \cite{1}.
As a potential element of nano- and microelectromechanical
devices, graphene can be separated by distances of the order of
tens or hundreds of nanometers from the other elements.
These are the distances at which the van der Waals and Casimir
forces caused by the zero-point and thermal fluctuations of the
electromagnetic field become dominant \cite{2}. That is why the
fluctuation induced {\it dispersion} forces from graphene have
attracted considerable attention in the last few years.

Many papers have been devoted to the calculation  of van der Waals
and Casimir forces between two graphene sheets
\cite{3,4,5,6,7,8,9,10,11}, a graphene sheet and a material plate
made of metallic, semiconductor or dielectric materials
\cite{4,5,6,7,8,9,12,13,14}, a graphene sheet and an atom,
a molecule, or other polarizable particle \cite{15,16,17,18}.
The calculations of the free energies and forces were performed
using the Lifshitz theory \cite{19} or its equivalent combined
with the reflection coefficients on graphene expressed via some
version of the density-density correlation function
\cite{5,6,8,9,18} or the polarization tensor for graphene defined
in (2+1)-dimensional space-time \cite{4,7,10,11,13,15,17}.
It was found \cite{5,7,10,11,13,17,19a} that the thermal Casimir
force in graphene systems is qualitatively different from the
case of plates made of conventional materials where the
classical regime holds at separations exceeding the so-called
{\it thermal length} equal to a few micrometers at room
temperature \cite{11,20}. However for the configuration of two
parallel graphene sheets, the classical behavior of the
Casimir interaction, characteristic for the case of large
separations (high temperatures), holds at separations
exceeding a few hundred nanometers \cite{5,7,10}, i.e.,
for an order of magnitude shorter separations than for
conventional materials.

Measurements of the Casimir force between two freestanding
graphene sheets
present additional difficulties, as compared to the case of
metallic or semiconductor surfaces (see Refs.~\cite{21,22,23}
for a review of experiments on measuring the Casimir force).
Because of this, in the pioneering experiment \cite{24} the
gradient of the Casimir force was measured between an
Au-coated sphere and a graphene sheet deposited on a
SiO${}_2$ film covering a Si plate. Measurements have been
performed by means of a dynamic atomic force microscope
(AFM) operated in the frequency-shift technique, i.e., using
a method well tested in previous experiments with metallic
test bodies \cite{25,26,27,28,29,30} and demonstrated its
high efficiency. However, the comparison of experiment with
theory in Ref.~\cite{24} used an approximate additive method
where the gradients of the Casimir force between a
Si-SiO${}_2$ system and an Au-coated sphere, and between a
graphene sheet described by the Dirac model and the same
sphere, were computed independently and then added.
Such an approximation was applied in the absence of exact
reflection coefficients for graphene deposited on a
substrate at nonzero temperature (the available reflection
coefficients \cite{9,31,32} were expressed in terms of the
density-density correlation function or, equivalently,
the conductivity of graphene whose explicit dependence on
the temperature remained unknown). As a result, the
theoretical force gradient computed using an assumption of
additivity overestimated the measured force gradient.
This was explained \cite{24} by the fact that an additive
method does not take into account the screening of the
SiO${}_2$ surface by the graphene layer. Thus, up to now
a complete quantitative theory explaining the
measurement data of Ref.~\cite{24} has been missing leading
to some uncertainty in the demonstrated influence of graphene
on the Casimir force.

In this paper, we express the reflection coefficients for
a graphene sheet deposited on a multilayered substrate made of
conventional materials via the components of the polarization
tensor defined at any nonzero temperature and the dielectric
permittivities of  substrates. The obtained reflection
coefficients coincide with those found recently by another
method \cite{33}. Then, we substitute the obtained reflection
coefficients in the Lifshitz theory and calculate the
gradient of the Casimir force in the experimental configuration
of Ref.~\cite{24} with no additional assumptions or fitting
parameters. We demonstrate that the experimental data are in
a very good agreement with theory within the limits of
experimental errors. Thus, the developed theory confirms
the demonstration of the Casimir force from graphene in
Ref.~\cite{24} and can be used for interpretation of future
experiments on measuring the Casimir interaction from
graphene deposited on multilayered material substrates.

The paper is organized as follows. In Sec.~II we express
the reflection coefficients from
graphene deposited on a thick material plate
(semispace) in terms of
 the components of the polarization tensor.
Section~III contains the generalization of these reflection
coefficients for graphene deposited on a multilayered
substrate  and the comparison between the
developed theory and the experimental data of Ref.~\cite{24}.
In Sec.~IV the reader will find our conclusions and discussion.

\section{Reflection coefficients from graphene on a substrate
in terms of the polarization tensor}

First, we consider a thin film spaced above a thick plate
(semispace) parallel to it in vacuum separated by a gap of
thickness $d$. Let the film be characterized by the
amplitude reflection coefficient $r_1$ and the transmission
coefficient $t_1$, and the plate be characterized by the
amplitude reflection coefficient $r_2$ (we are interested in the
reflection coefficients calculated along the imaginary
frequency axis $\omega=i\xi$).
Note that the amplitude coefficients correspond to ratios of the
reflected (transmitted) complex-valued amplitudes of the electric
field to that of the incident field. They are different from the
power coefficients which are the fractions of the incident power
that is reflected or refracted at the interface.
Then, taking into account multiple
reflections on the plane of the film and on the upper
boundary plane of the plate, one obtains the reflection
coefficient $R$ on the system consisting of the film, the gap
and the plate (see, for instance, Ref.~\cite{34})
\begin{eqnarray}
R&=&r_1+t_1r_2t_1e^{-2dq}\sum_{n=0}^{\infty}
\left( r_1r_2e^{-2dq}\right)^n
\nonumber \\
&=&
r_1+\frac{t_1r_2t_1e^{-2dq}}{1-r_1r_2e^{-2dq}},
\label{eq1}
\end{eqnarray}
\noindent
where
\begin{equation}
q=\left(k_{\bot}^2+\frac{\xi^2}{c^2}\right)^{1/2}
\label{eq2}
\end{equation}
\noindent
and $k_{\bot}$ is the projection of the wave vector on the
plane of the film. Note that Eq.~(\ref{eq1}) is applicable
to both the transverse magnetic (TM) and transverse electric
(TE) reflection coefficients defined for the two independent
polarizations of the electromagnetic field.

Now we apply Eq.~(\ref{eq1}) to a graphene sheet $(g)$ spaced
at a height $d$ above a material semispace $(s)$ characterized
by the frequency-dependent dielectric permittivity
$\varepsilon_1(i\xi)$. We consider in succession the cases of
TM and TE reflection coefficients.
Although for the TM polarization of the electromagnetic field
there is no general relationship between the amplitude
reflection and transmission coefficients \cite{34a}, for a
graphene sheet in vacuum it can be shown that \cite{32}
\begin{equation}
t_{\rm TM}^{(g)}=1-r_{\rm TM}^{(g)}.
\label{eq3}
\end{equation}
\noindent
Then, by putting $r_1=r_{\rm TM}^{(g)}$, $t_1=t_{\rm TM}^{(g)}$
and $r_2=r_{\rm TM}^{(s)}$ in Eq.~(\ref{eq1}), and using
Eq.~(\ref{eq3}), one obtains
\begin{eqnarray}
R_{\rm TM}^{(g,s)}&=&r_{\rm TM}^{(g)}
+\frac{t_{\rm TM}^{(g)}r_{\rm TM}^{(s)}t_{\rm TM}^{(g)}e^{-2dq}}{1-
r_{\rm TM}^{(g)}r_{\rm TM}^{(s)}e^{-2dq}}
\nonumber \\
&=&
\frac{r_{\rm TM}^{(g)}+r_{\rm TM}^{(s)}\left(1-
2r_{\rm TM}^{(g)}\right)e^{-2dq}}{1-
r_{\rm TM}^{(g)}r_{\rm TM}^{(s)}e^{-2dq}}.
\label{eq4}
\end{eqnarray}

In order to obtain the reflection coefficient  from graphene
deposited on a thick plate (semispace), we put $d\to 0$
in Eq.~(\ref{eq4}) and arrive at
\begin{equation}
R_{\rm TM}^{(g,s)}=
\frac{r_{\rm TM}^{(g)}+r_{\rm TM}^{(s)}\left(1-
2r_{\rm TM}^{(g)}\right)}{1-
r_{\rm TM}^{(g)}r_{\rm TM}^{(s)}}.
\label{eq5}
\end{equation}
\noindent
Note that Eq.~(\ref{eq1}) was used in Ref.~\cite{18} for
application to the TM mode of the electromagnetic field, but
with an incorrect relationship
$t_{\rm TM}^{(g)}=1+r_{\rm TM}^{(g)}$ instead of Eq.~(\ref{eq3}).

For a graphene sheet in vacuum the TM reflection coefficient in
terms of the polarization tensor takes the form \cite{7}
\begin{equation}
r_{\rm TM}^{(g)}\equiv r_{\rm TM}^{(g)}(i\xi,k_{\bot})=
\frac{q\Pi_{00}(i\xi,k_{\bot})}{q\Pi_{00}(i\xi,k_{\bot})+
2\hbar k_{\bot}^2},
\label{eq6}
\end{equation}
\noindent
where an exact expression for the 00-component of the
polarization tensor $\Pi_{00}$ in (2+1)-dimensional
space-time can be found in
Refs.~\cite{7,10,13,17} (see Sec.~III for the expression
used in computations). Note that $\Pi_{00}$ depends on
the temperature as a parameter.

The TM reflection coefficient on the boundary between a vacuum
and a semispace is the well known Fresnel coefficient \cite{2,19}
\begin{equation}
r_{\rm TM}^{(s)}\equiv r_{\rm TM}^{(s)}(i\xi,k_{\bot})=
\frac{\varepsilon_1(i\xi)q-k_1}{\varepsilon_1(i\xi)q+k_1},
\label{eq7}
\end{equation}
\noindent
where
\begin{equation}
k_1\equiv k_1(i\xi,k_{\bot})=\left[k_{\bot}^2+
\varepsilon_1(i\xi)\frac{\xi^2}{c^2}\right]^{1/2}.
\label{eq8}
\end{equation}
\noindent
Substituting Eq.~(\ref{eq7}) in
Eq.~(\ref{eq5}), we arrive at the TM reflection coefficient
from the graphene sheet deposited on a material semispace
\begin{equation}
R_{\rm TM}^{(g,s)}(i\xi,k_{\bot})=
\frac{\varepsilon_1q+k_1\left(\frac{q}{\hbar k_{\bot}^2}
\Pi_{00}-1\right)}{\varepsilon_1q+k_1\left(\frac{q}{\hbar k_{\bot}^2}
\Pi_{00}+1\right)},
\label{eq9}
\end{equation}
\noindent
where $\Pi_{00}\equiv \Pi_{00}(i\xi,k_{\bot})$ and
$\varepsilon_1\equiv\varepsilon_1(i\xi)$.

For convenience in computations, we express Eq.~(\ref{eq9}) in
terms of the dimensionless variables
\begin{equation}
y=2aq, \qquad
\zeta=2a\xi/c
\label{eq10}
\end{equation}
\noindent
leading to
\begin{equation}
k_1=\frac{1}{2a}\left[y^2+
(\varepsilon_1-1)\zeta^2\right]^{1/2}.
\label{eq10a}
\end{equation}
\noindent
The reflection coefficient (\ref{eq9}) in terms of the new variables
takes the form
\begin{equation}
R_{\rm TM}^{(g,s)}(i\zeta,y)=
\frac{\varepsilon_1y+\sqrt{y^2+(\varepsilon_1-1)\zeta^2}
\left(\frac{y}{y^2-\zeta^2}\tilde{\Pi}_{00}-
1\right)}{\varepsilon_1y+\sqrt{y^2+(\varepsilon_1-1)\zeta^2}
\left(\frac{y}{y^2-\zeta^2}\tilde{\Pi}_{00}+
1\right)},
\label{eq11}
\end{equation}
\noindent
where
\begin{equation}
\tilde{\Pi}_{00}\equiv \tilde{\Pi}_{00}(i\zeta,y)
=\frac{2a}{\hbar}\Pi_{00}
\label{eq12}
\end{equation}
\noindent
is the dimensionless polarization tensor \cite{10,13,17}.

We now proceed with the case of the TE reflection coefficient
for a graphene sheet at a height $d$ in vacuum above a
semispace. There is a general relationship between the amplitude
reflection and transmission coefficients in the case of TE
polarization of the electromagnetic field \cite{34a}
\begin{equation}
t_{\rm TE}^{(g)}=1+r_{\rm TE}^{(g)}.
\label{eq13}
\end{equation}
\noindent
Substituting this in Eq.~(\ref{eq1}) and
 putting $r_1=r_{\rm TE}^{(g)}$, $t_1=t_{\rm TE}^{(g)}$
and $r_2=r_{\rm TE}^{(s)}$, we obtain
\begin{eqnarray}
R_{\rm TE}^{(g,s)}&=&r_{\rm TE}^{(g)}
+\frac{t_{\rm TE}^{(g)}r_{\rm TE}^{(s)}t_{\rm TE}^{(g)}e^{-2dq}}{1-
r_{\rm TE}^{(g)}r_{\rm TM}^{(s)}e^{-2dq}}
\nonumber \\
&=&
\frac{r_{\rm TE}^{(g)}+r_{\rm TE}^{(s)}\left(1+
2r_{\rm TE}^{(g)}\right)e^{-2dq}}{1-
r_{\rm TE}^{(g)}r_{\rm TE}^{(s)}e^{-2dq}}.
\label{eq14}
\end{eqnarray}

In the limiting case $d\to 0$ one obtains from Eq.~(\ref{eq14})
the TE reflection coefficient for a graphene deposited on a
semispace
\begin{equation}
R_{\rm TE}^{(g,s)}=
\frac{r_{\rm TE}^{(g)}+r_{\rm TE}^{(s)}\left(1+
2r_{\rm TE}^{(g)}\right)}{1-
r_{\rm TE}^{(g)}r_{\rm TE}^{(s)}}.
\label{eq15}
\end{equation}

The TE reflection coefficient for a graphene sheet in vacuum is
given by \cite{7}
\begin{equation}
r_{\rm TE}^{(g)}\equiv r_{\rm TE}^{(g)}(i\xi,k_{\bot})=
-\frac{k_{\bot}^2\Pi_{\rm tr}(i\xi,k_{\bot})-
q^2\Pi_{00}(i\xi,k_{\bot})}{k_{\bot}^2\Pi_{\rm tr}(i\xi,k_{\bot})-
q^2\Pi_{00}(i\xi,k_{\bot})+
2\hbar k_{\bot}^2q},
\label{eq16}
\end{equation}
\noindent
where
the exact expression for  $\Pi_{\rm tr}$ can be found in
Refs.~\cite{7,10,13,17} (see also Sec.~III).

The Fresnel TE reflection coefficient from a semispace
to a vacuum is given by \cite{2,19}
\begin{equation}
r_{\rm TE}^{(s)}\equiv r_{\rm TE}^{(s)}(i\xi,k_{\bot})=
\frac{q-k_1}{q+k_1}.
\label{eq17}
\end{equation}
\noindent
Substituting Eqs.~(\ref{eq16}) and (\ref{eq17}) in
Eq.~(\ref{eq15}), one obtains the TE reflection coefficient
from the graphene sheet deposited on a  semispace
\begin{equation}
R_{\rm TE}^{(g,s)}(i\xi,k_{\bot})=
\frac{q-k_1-\frac{1}{\hbar k_{\bot}^2}\left(
k_{\bot}^2\Pi_{\rm tr}-q^2\Pi_{00}\right)}{q+k_1+
\frac{1}{\hbar k_{\bot}^2}\left(
k_{\bot}^2\Pi_{\rm tr}-q^2\Pi_{00}\right)},
\label{eq18}
\end{equation}
\noindent
where $\Pi_{\rm tr}\equiv \Pi_{\rm tr}(i\xi,k_{\bot})$.
If we take into account the connections between the
polarization tensor and the
temperature-dependent density-density correlation
functions and nonlocal dielectric permittivities
found in Ref.~\cite{33}, it can be seen
that the reflection coefficients
(\ref{eq9}) and (\ref{eq18}) coincide with the respective
coefficients derived in Refs.~\cite{9,31,32} from the
exact electrodynamic boundary conditions.

In terms of dimensionless variables (\ref{eq10})
the reflection coefficient (\ref{eq18})
takes the form
\begin{equation}
R_{\rm TE}^{(g,s)}(i\zeta,y)=
\frac{y-\sqrt{y^2+(\varepsilon_1-1)\zeta^2}-\left(\tilde{\Pi}_{\rm tr}
-\frac{y^2}{y^2-\zeta^2}\tilde{\Pi}_{00}\right)}{y+
\sqrt{y^2+(\varepsilon_1-1)\zeta^2}+\left(\tilde{\Pi}_{\rm tr}
-\frac{y^2}{y^2-\zeta^2}\tilde{\Pi}_{00}\right)},
\label{eq19}
\end{equation}
\noindent
where
\begin{equation}
\tilde{\Pi}_{\rm tr}\equiv \tilde{\Pi}_{\rm tr}(i\zeta,y)
=\frac{2a}{\hbar}\Pi_{\rm tr}.
\label{eq20}
\end{equation}
\noindent
The representation (\ref{eq19}) is convenient for use in
numerical computations.

\section{Comparison between experiment and theory}

In the experiment of Ref.~\cite{24} the gradient of the
Casimir force was measured between an Au-coated hollow glass
sphere of radius $R=54.1\,\mu$m and a large area graphene
sheet deposited on a $D=300\,$nm thick SiO${}_2$ film
covering a B-doped Si plate of $500\,\mu$m thickness.
The thickness of the Au coating on the sphere was measured
to be 280\,nm.
With respect to the Casimir force, the sphere can be considered
as completely gold and the Si plate as an infinitely thick semispace
\cite{2}. However, the exact thickness of the SiO${}_2$ film
should be taken into account in computations of the theoretical
Casimir force.

In the dynamic measurement scheme by means of the  AFM the sphere was
attached to a cantilever oscillating with the natural resonant
frequency $\omega_0$. Under the influence of the external force,
electric and Casimir, the resonant frequency was modified.
The change in the frequency $\Delta\omega=\omega_r-\omega_0$,
where $\omega_r$ is the resonance frequency in the presence of
external force, was measured by means of a phase-locked loop and
recorded as a function of separation $a$ between the sphere and
graphene surfaces. This frequency change is proportional to the
gradient of the external force.

Measurements were performed in high vacuum down to
$10^{-9}\,$Torr with two graphene samples in two different ways.
In the first case, after the electrostatic calibration (i.e.,
determination of the calibration constant, residual potential
difference, and the separation at the closest approach between
the test bodies by applying different voltages), the gradient
of the Casimir force was subtracted from the total measured
force gradient resulting in the gradient of the Casimir force.
In the second case, only the compensating voltage equal to the
residual potential difference was applied to graphene, and
the gradient of the Casimir force was an immediately measured
quantity.

Altogether 84 force-distance relations have been measured
(20 with different applied voltages and 22 with applied
compensating voltage for each of the two graphene samples).
All the results were found to be in a very good mutual agreement
in the limits of the experimental errors \cite{24}.
Below we perform the theory-experiment comparison for the two sets
of mean measured gradients of the Casimir force obtained for
the two
graphene samples with applied compensating voltages.
In this case the total error of the measured gradients of the
Casimir force is slightly smaller than for the other two sets
with subtracted electrostatic forces because the latter are
calculated with some errors which should be added to the total
experimental error common to both ways of measurement.

\begin{figure}[b]
\vspace*{-5cm}
\centerline{\hspace*{1cm}
\includegraphics{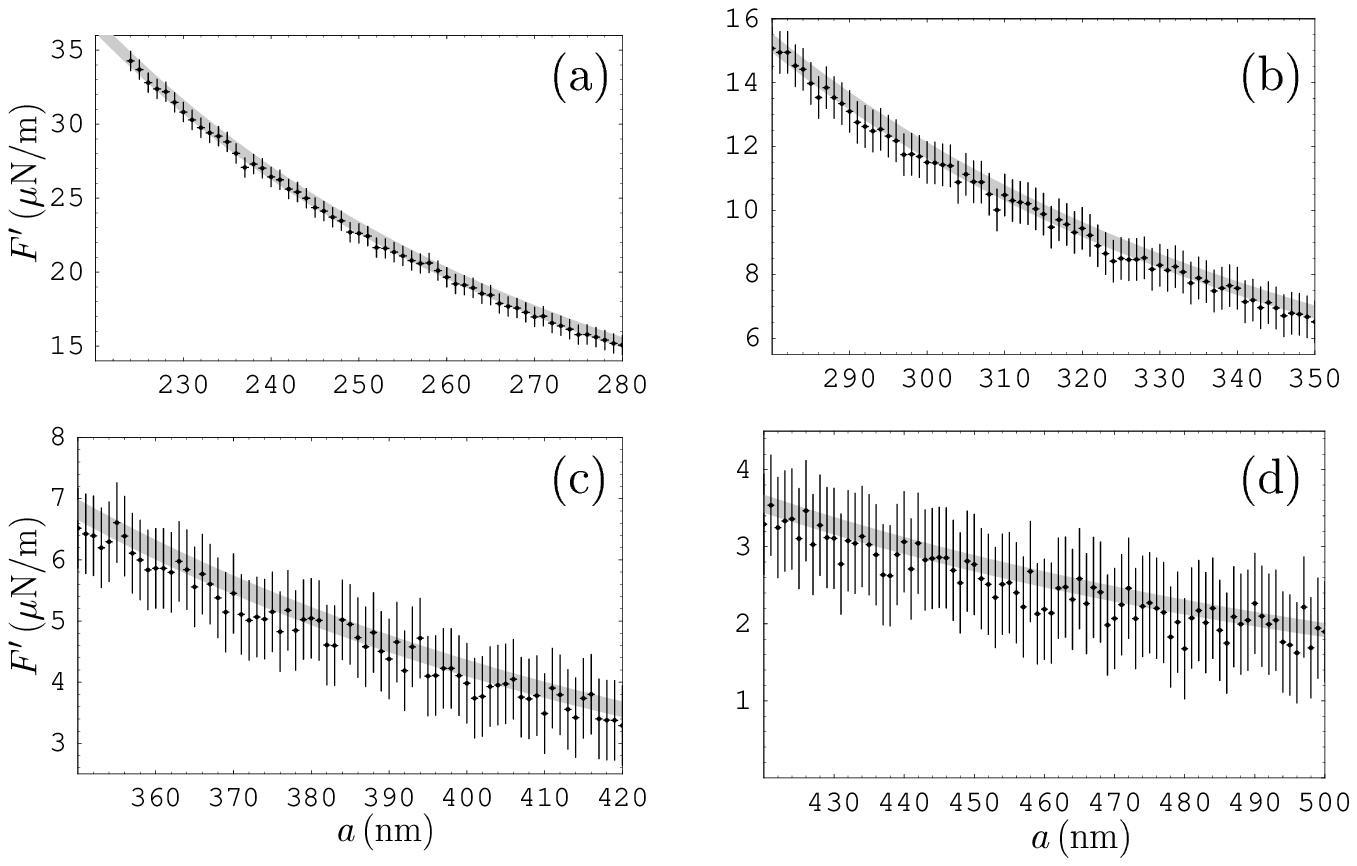}
}
\vspace*{-11cm}
\caption{\label{fg1}
The experimental data for
the gradient of the Casimir force
between an Au-coated sphere and graphene deposited on a
SiO${}_2$ film covering a Si plate (the first sample)
are shown as crosses plotted at a 67\% confidence level
over different separation regions. The gray bands present the
theoretical force gradients computed using the exact
reflection coefficients for graphene on a
multilayered substrate derived
here in terms of the polarization tensor.
}
\end{figure}

\begin{figure}[b]
\vspace*{-5cm}
\centerline{\hspace*{1cm}
\includegraphics{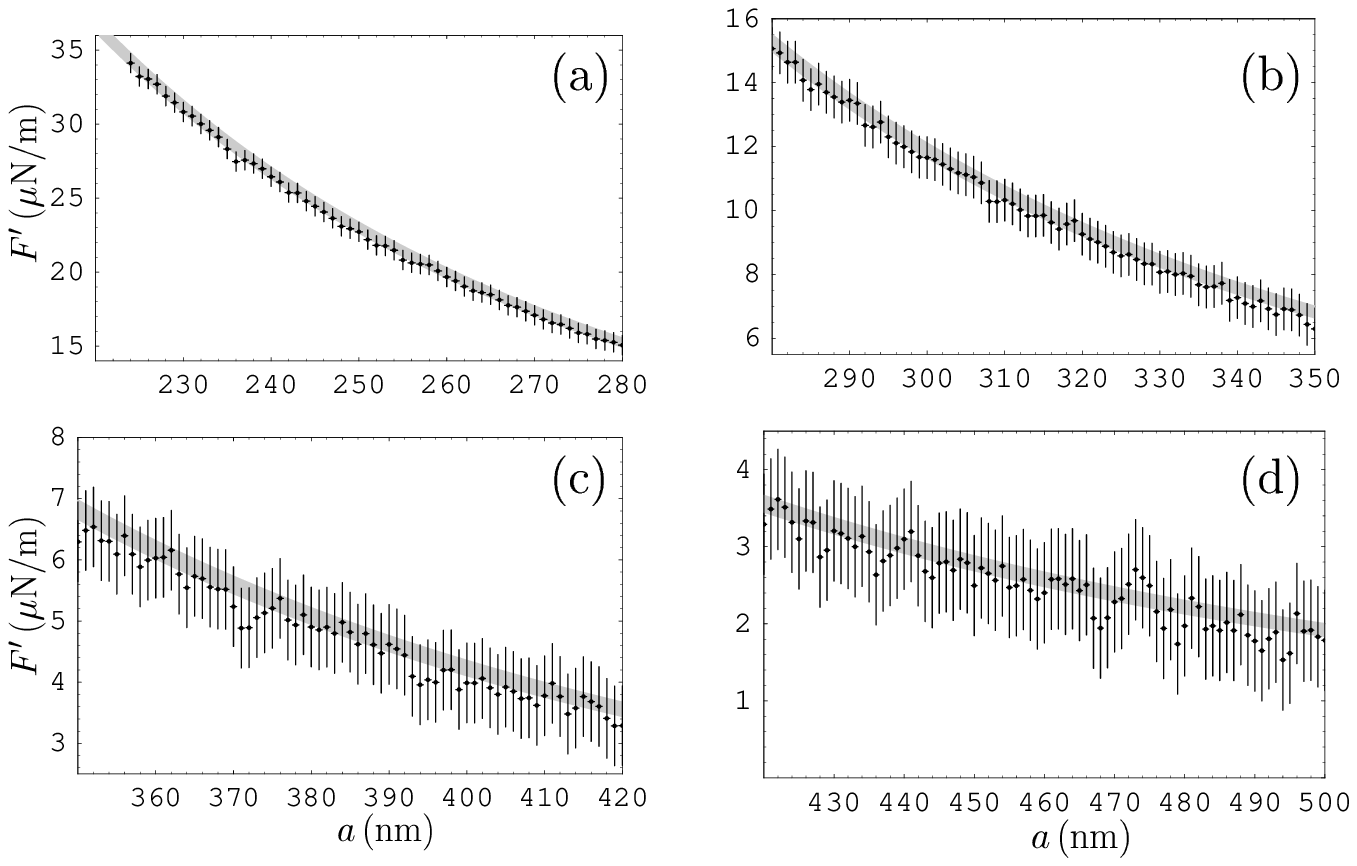}
}
\vspace*{-11cm}
\caption{\label{fg2}
The experimental data for
the gradient of the Casimir force
between an Au-coated sphere and graphene deposited on a
SiO${}_2$ film covering a Si plate (the second sample)
are shown as crosses plotted at a 67\% confidence level
over different separation regions. The gray bands present the
theoretical force gradients computed using the exact
reflection coefficients for graphene on a
multilayered substrate derived
here in terms of the polarization tensor.
}
\end{figure}

In Figs.~\ref{fg1} and \ref{fg2} the mean gradients of the
Casimir force, measured \cite{24} for the first and second
graphene samples, are indicated as crosses. The averaging was
performed over 22 force-distance relations obtained for each of
the two samples.
The horizontal arms of the crosses indicate twice the error
$\Delta a=0.4\,$nm in measurements of absolute separations
between the surfaces. The vertical arms are twice the error
$\Delta F^{\prime}=0.64\,\mu$N/m in measurements of the
gradient of the Casimir force.
All errors are indicated at the 67\% confidence level.
Measurements were performed over
the separation region from 224 to 500\,nm.

The gradient of the Casimir force between an Au sphere and
graphene sheet deposited on a SiO${}_2$ film covering a Si plate
(semispace) was calculated using the Lifshitz formula in the
proximity force approximation \cite{2}. For convenience in
computations, we use the dimensionless variables (\ref{eq10})
and obtain
\begin{eqnarray}
&&
F^{\prime}(a)=\frac{k_BTR}{4a^3}\sum_{l=0}^{\infty}\!
{\vphantom{\sum}}^{\prime}\int_{\zeta_l}^{\infty}\!\!\! y^2dy
\left[\frac{r_{\rm TM}^{(\rm Au)}(i\zeta_l,y)
R_{\rm TM}^{(g,f,s)}(i\zeta_l,y)}{e^y-
r_{\rm TM}^{(\rm Au)}(i\zeta_l,y)
R_{\rm TM}^{(g,f,s)}(i\zeta_l,y)}\right.
\nonumber \\
&&~~~~
\left.+
\frac{r_{\rm TE}^{(\rm Au)}(i\zeta_l,y)
R_{\rm TE}^{(g,f,s)}(i\zeta_l,y)}{e^y-
r_{\rm TE}^{(\rm Au)}(i\zeta_l,y)
R_{\rm TE}^{(g,f,s)}(i\zeta_l,y)}\right],
\label{eq21}
\end{eqnarray}
\noindent
where $k_B$ is the Boltzmann constant, $T=300\,$K is the
temperature at the laboratory,
and the dimensionless quantities $\zeta_l=2a\xi_l/c$ are
expressed via the Matsubara frequencies
$\xi_l=2\pi k_BTl/\hbar$ with $l=0,\,1,\,2,\,\ldots\,\,$.
Note that under the condition $a\ll R$, which is satisfied
in our case with a wide safety margin, the corrections to
PFA in sphere-plate geometry are negligibly small
\cite{35,36,37,38}. Now we specify the reflection coefficients
$r_{\rm TM,TE}^{(\rm Au)}$ and $R_{\rm TM,TE}^{(g,f,s)}$
entering Eq.~(\ref{eq21}).

The first of them is the standard amplitude Fresnel coefficient
on the Au semispace given by Eqs.~(\ref{eq7}) and (\ref{eq17}).
In terms of dimensionless variables, for the TM and TE
polarizations, it is given by
\begin{eqnarray}
&&
r_{\rm TM}^{(\rm Au)}(i\zeta_l,y)=
\frac{\varepsilon_l^{\,(\rm Au)}y-
\sqrt{y^2+(\varepsilon_l^{\,(\rm Au)}-1)\zeta_l^2}}{\varepsilon_l^{\,(\rm Au)}
y+\sqrt{y^2+(\varepsilon_l^{\,(\rm Au)}-1)\zeta_l^2}},
\nonumber \\
&&
r_{\rm TE}^{(\rm Au)}(i\zeta_l,y)=
\frac{y-
\sqrt{y^2+(\varepsilon_l^{\,(\rm Au)}-1)\zeta_l^2}}{y+
\sqrt{y^2+(\varepsilon_l^{\,(\rm Au)}-1)\zeta_l^2}},
\label{eq22}
\end{eqnarray}
\noindent
where
$\varepsilon_l^{\,(\rm Au)}\equiv\varepsilon^{\,(\rm Au)}(ic\zeta_l/2a)$.
The latter quantity is found (see reviews \cite{2,21} and
references therein) by means of the Kramers-Kronig relation from
the optical data for ${\rm Im}\varepsilon^{\,(\rm Au)}$ given
over a wide frequency range in Ref.~\cite{39} and extrapolated to
zero frequency. It is well known \cite{2,21} that there are two
approaches to this extrapolation using the Drude and the plasma
models leading to different results for the thermal Casimir
force between two metallic surfaces. For a metallic surface
interacting with graphene the differences arising from the use of
two approaches are, however, negligibly
small \cite{13,24}
and are included here into the magnitude of the theoretical error.

The reflection coefficients from a graphene sheet deposited
on a SiO${}_2$ film covering a Si plate,
$R_{\rm TM,TE}^{(g,f,s)}$, can be written using the formalism,
developed in Sec.~II and the standard formulas of the Lifshitz
theory describing the reflection coefficients from planar
layered structures \cite{2,40,41}
\begin{equation}
R_{\rm TM,TE}^{(g,f,s)}(i\zeta_l,y)=
\frac{R_{\rm TM,TE}^{(g,s)}(i\zeta_l,y)+
r_{\rm TM,TE}^{(f,s)}(i\zeta_l,y)e^{-2Dk_1}}{1+
R_{\rm TM,TE}^{(g,s)}(i\zeta_l,y)
r_{\rm TM,TE}^{(f,s)}(i\zeta_l,y)e^{-2Dk_1}},
\label{eq23}
\end{equation}
\noindent
where $k_1$ is defined in Eq.~(\ref{eq10a}).
The reflection coefficients $R_{\rm TM,TE}^{(g,s)}$
are given in Eqs.~(\ref{eq11}) and (\ref{eq19}),
respectively. They describe the reflection from a
graphene sheet deposited on a SiO${}_2$ semispace, and,
thus,
$\varepsilon_{1}=\varepsilon^{\,({\rm SiO}_2)}(ic\zeta_l/2a)
\equiv\varepsilon_{1l}$.
The reflection coefficients $r_{\rm TM,TE}^{(f,s)}$ describe
the reflection on the boundary plane between the two
semispaces made of SiO${}_2$ and Si.
These are the standard, Fresnel, reflection coefficients.
In terms of dimensionless variables they are given by
\begin{eqnarray}
&&
r_{\rm TM}^{(f,s)}(i\zeta_l,y)=
\frac{\varepsilon_{2l}k_1-
\varepsilon_{1l}k_2}{\varepsilon_{2l}k_1+
\varepsilon_{1l}k_2}
\nonumber \\
&&
r_{\rm TE}^{(f,s)}(i\zeta_l,y)=
\frac{k_1-k_2}{k_1+k_2},
\label{eq24}
\end{eqnarray}
\noindent
where
$\varepsilon_{2l}\equiv\varepsilon^{\,(\rm Si)}(ic\zeta_l/2a)$
and, similar to Eq.~(\ref{eq10a}),
\begin{equation}
k_2=\frac{1}{2a}\left[y^2+
(\varepsilon_{2l}-1)\zeta_l^2\right]^{1/2}.
\label{eq25}
\end{equation}

Now we discuss explicit expressions for all the quantities entering
Eq.~(\ref{eq23}). According to Eqs.~(\ref{eq11}) and (\ref{eq19}),
the coefficient $R_{\rm TM,TE}^{(g,s)}$ depends on the
components of the polarization tensor. As was shown in
Refs.~\cite{7,10,13,17}, an explicit dependence of the polarization
tensor on $T$ influences the computational results only through
the contribution from the zero Matsubara frequency $\zeta_0=0$,
whereas all contributions with $l\geq 1$ can be calculated with
the polarization tensor defined at $T=0$. Because of this, in
computations below we use the following temperature-dependent
expressions at $\zeta_0=0$ entering Eqs.~(\ref{eq11}) and
(\ref{eq19}) \cite{7,13}
\begin{eqnarray}
&&
\tilde{\Pi}_{00}(0,y)=\frac{8\alpha}{\tilde{v}_F^2}\left[
\frac{\tau}{\pi}\int_{0}^{1}\!\!\! dx\ln\left(2
\cosh\frac{\pi\theta}{\tau}\right)\right.
\nonumber \\
&&~~~~~~\left.
-\tilde{\Delta}^2\int_{0}^{1}\!\frac{dx}{\theta}
\tanh\frac{\pi\theta}{\tau}\right],
\label{eq26} \\
&&
\tilde{\Pi}_{\rm tr}(0,y)-\tilde{\Pi}_{00}(0,y)=
8\alpha\tilde{v}_F^2y^2\int_{0}^{1}\!\!dx\frac{x(1-x)}{\theta}
\tanh\frac{\pi\theta}{\tau}.
\nonumber
\end{eqnarray}
\noindent
Here, $\alpha=e^2/(\hbar c)$ is the fine-structure constant,
$v_F\approx 9\times 10^5\,$m/s is the Fermi velocity in
graphene \cite{42,43}, $\tilde v_F\equiv v_F/c$,
the temperature parameter is $\tau=4\pi ak_BT/(\hbar c)$,
and the following notation is introduced
\begin{equation}
\theta\equiv\theta(x,y)=\left[\tilde{\Delta}^2+
x(1-x)\tilde{v}_F^2y^2\right]^{1/2},
\label{eq27}
\end{equation}
\noindent
where $\Delta$ is gap parameter of graphene and
$\tilde{\Delta}\equiv 2a\Delta/(\hbar c)$.
The gap parameter takes into account that the Dirac-type
excitations in graphene become massive under the influence
of electron-electron interaction, substrates, and defects
of the structure \cite{43a,43b,43c,43d,43e}.
The exact value of $\Delta$ is unknown but its maximum
value is estimated as 0.1\,eV \cite{4}.

At all nonzero Matsubara frequencies one can
use in  Eqs.~(\ref{eq11}) and (\ref{eq19})
the following expressions found \cite{4,7,13} at $T=0$:
\begin{eqnarray}
&&
\tilde{\Pi}_{00}(i\zeta_l,y)=\alpha
\frac{y^2-\zeta_l^2}{f^2(\zeta_l,y)}\,\Phi(\zeta_l,y)
\label{eq28} \\
&&
\tilde{\Pi}_{\rm tr}(i\zeta_l,y)-
\frac{y^2}{y^2-\zeta_l^2}\tilde{\Pi}_{00}(i\zeta_l,y)=
\alpha\Phi(\zeta_l,y),
\nonumber
\end{eqnarray}
\noindent
where the two notations are introduced
\begin{eqnarray}
&&
f(\zeta_l,y)=\left[\tilde{v}_F^2y^2+(1-\tilde{v}_F^2)
\zeta_l^2\right]^{1/2},
\label{eq29}\\
&&
\Phi(\zeta_l,y)=4\tilde{\Delta}+2f(\zeta_l,y)\left[
1-\frac{4\tilde{\Delta}^2}{f^2(\zeta_l,y)}\right]\,
\arctan\frac{f(\zeta_l,y)}{2\tilde{\Delta}}.
\nonumber
\end{eqnarray}

Two more quantities, which are needed to calculate the
reflection coefficients (\ref{eq24}), are the dielectric
permittivities of Si and SiO${}_2$ at the imaginary
Matsubara frequencies.
The Si plate used in Ref.~\cite{24} had a resistivity
between 0.001 and $0.005\,\Omega\,$cm, which corresponds
\cite{44} to a charge carrier density
$n\approx(1.6\div 7.8)\times 10^{19}\,\mbox{cm}^{-3}$.
For B-doped Si the dielectric-to-metal transition occurs
\cite{45} at $n_c\approx 3.95\times 10^{18}\,\mbox{cm}^{-3}$.
Thus, the Si used was of metal-type with the plasma frequency
$\omega_p$ between
$5\times 10^{14}\,$rad/s and $11\times 10^{14}\,$rad/s
 \cite{46} and
 $\gamma\approx 1.1\times 10^{14}\,$rad/s for
the relaxation parameter \cite{24}.
In our computations we used
$\varepsilon_{2l}=\varepsilon^{\,(\rm Si)}(i\xi_l)$ obtained
\cite{47} by means of the Kramers-Kronig relation from
the optical data \cite{48} extrapolated to zero frequency
either by the Drude or by the plasma model. Similar to the
case of Au interacting with graphene, here different types
of extrapolation lead to only a minor differences in the
resulting force gradients which are included in the
theoretical error. As to the dielectric permittivity
$\varepsilon_{1l}=\varepsilon^{\,({\rm SiO}_2)}(i\xi_l)$,
a sufficiently accurate expression for it presented in
Ref.~\cite{49} was used in computations.

Finally, the gradients of the Casimir force in the
experimental configuration of Ref.~\cite{24} were
computed by Eq.~(\ref{eq21}) with the reflection coefficients
presented in Eqs.~(\ref{eq22}) and (\ref{eq23}).
The obtained force gradients were corrected for the presence
of surface roughness on both surfaces. For this purpose the
rms roughness was measured by means of AFM and found to be
equal to 1.6 and 1.5\,nm on the sphere and graphene,
respectively. It was taken into account using the
multiplicative approach \cite{2,21} which is sufficiently
precise for so small roughness at relatively large
separations above 200\,nm. It was shown that maximum
contribution of roughness to the force gradient achieved at
the shortest separation of $a=224\,$nm is equal to only
0.1\% of the calculated results.

The computed gradients of the Casimir force are shown as
gray bands in Figs.~\ref{fg1}(a-d) and \ref{fg2}(a-d)
in comparison with the experimental data obtained for the
first and second graphene samples, respectively.
The widths of the bands are determined by the uncertainty
in the value of $\omega_p$ for a Si plate within the interval
indicated above, differences between the predictions of the
Drude and plasma model extrapolations of the optical data for
Au and Si, and by the uncertainty of the mass gap parameter of
graphene within the interval from 0 to 0.1\,eV.
As can be seen in Figs.~\ref{fg1} and \ref{fg2}, our theory
describing the reflection coefficients from graphene deposited
on a substrate in terms of the polarization tensor is in
a very good agreement with the measurement data.

\begin{figure}[b]
\vspace*{-12cm}
\centerline{\hspace*{3cm}
\includegraphics{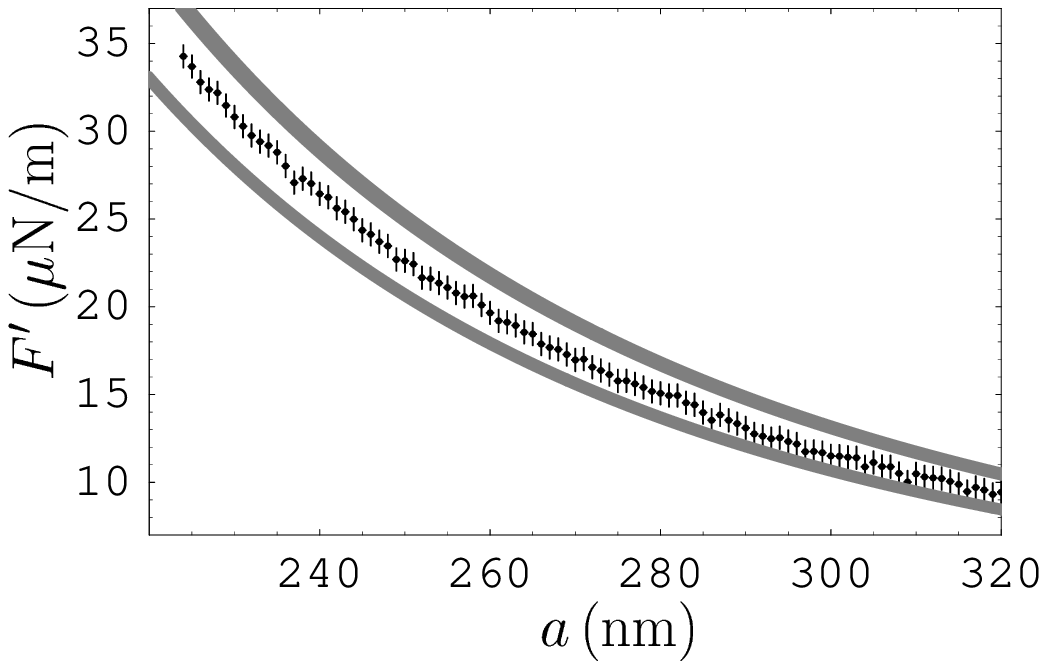}
}
\vspace*{-6cm}
\caption{\label{fg3}
The experimental data for
the gradient of the Casimir force
between an Au-coated sphere and graphene deposited on a
SiO${}_2$ film covering a Si plate (the first sample)
are shown as crosses plotted at a 67\% confidence level
over the separation region below 320\,nm.
The gray bands present the
theoretical force gradients between an Au-coated
sphere and a substrate consisting of a SiO${}_2$ film
covering a Si plate computed using
the standard Lifshitz theory (the lower band)
and between an Au-coated sphere and
graphene deposited on this substrate  using
the additive approach (the upper band).
}
\end{figure}

To make the advantages of the suggested theory more transparent,
we again present in Fig.~\ref{fg3} the experimental data for
$F^{\prime}$ obtained with the first graphene sample in
comparison with the two gray theoretical bands. The lower band
shows the gradient of the Casimir force between an Au-coated
sphere and a substrate consisting of a SiO${}_2$ film covering
a Si plate with no graphene coating calculated
using the standard Lifshitz theory (measurement of this force
gradient presents difficulties due to electric charges localized
on the dielectric surface).
The upper band shows the
force gradient between an Au-coated sphere and graphene deposited
on this substrate obtained using an additive approach (i.e., computed by
adding the force gradient from Au-graphene interaction to the
lower band). As can be seen from the figure, the lower band
underestimates whereas the upper band overestimates the
measured force gradients. The latter was explained in
Ref.~\cite{24} by the fact that the additive approach does not
take into account the screening of the SiO${}_2$ film by the
graphene sheet. The theory developed here takes the effects of
nonadditivity into account and brings theoretical predictions in
agreement with the measurement data.

\section{Conclusions and discussion}

In the foregoing, we have developed a theory of the Casimir
interaction for a graphene deposited on multilayered substrate
made of ordinary materials. The reflection coefficients on
substrates coated with graphene were expressed via
components of the polarization tensor of graphene in
(2+1)-dimensional space-time and dielectric permittivities of
substrate materials in the imaginary Matsubara frequencies.
The suggested theory allows calculation of the Casimir
interaction between two graphene-coated multilayered structures
and between the test body made of an ordinary material  and
the graphene-coated multilayered substrate at any temperature.
It allows generalization for the case of doped graphene sheets.

The developed theory was applied to the configuration of the
experiment \cite{24} on measuring the gradient of the Casimir
force between an Au-coated sphere and a graphene sheet deposited
on a SiO${}_2$ film covering a Si plate. Previously the measurement
data of this experiment was compared with only an approximate
additive theory and agreement with the computational results
was not achieved. We have performed computations of the gradient
of the Casimir force using the reflection coefficients on a
multilayered substrate coated with graphene which are expressed
via the polarization tensor. Good agreement between the
new theory and the measurement data was demonstrated with no
fitting parameters in the limits of the experimental
 errors and uncertainties.

The achieved agreement between the experimental data and the
complete theory applicable to a graphene deposited on
substrates allows to conclude with certainty that the
experiment of Ref.~\cite{24} demonstrates influence of graphene
sheet on the Casimir interaction. This conclusion opens
prospective opportunities for tailoring the Casimir force in
nanostructures by using graphene.

\section*{Acknowledgments}

This work was supported by the DOE Grant
No.\ DEF010204ER46131 (U.M.).
The authors are grateful to Bo~E.~Sernelius for stimulating
discussions.


\end{document}